\documentclass[12pt,preprint]{aastex}

\shorttitle{INTEGRAL/IBIS Extragalactic survey}
\shortauthors{Bassani et al.}
\begin{document}

\title{INTEGRAL/IBIS Extragalactic survey: 20-100 keV selected AGN\altaffilmark{1}}

\author{L.~Bassani\altaffilmark{2}, M.~Molina\altaffilmark{3}, A.~Malizia\altaffilmark{2},
J.B.~Stephen\altaffilmark{2}, A.J.~Bird\altaffilmark{3}, A.~Bazzano\altaffilmark{4},
G.~B\'elanger\altaffilmark{5}, A.J.~Dean\altaffilmark{3}, A.~De~Rosa\altaffilmark{4},
P.~Laurent\altaffilmark{5}, F.~Lebrun\altaffilmark{5}, P.~Ubertini\altaffilmark{4},
and R.~Walter\altaffilmark{6}}

\altaffiltext{1}{Based on observations obtained with the ESA science mission {\it INTEGRAL}}
\altaffiltext{2}{IASF-Bologna/INAF, via P. Gobetti 101, 40129 Bologna, Italy; bassani@iasfbo.inaf.it, malizia@iasfbo.inaf.it,
stephen@iasfbo.inaf.it}
\altaffiltext{3}{School of Physics and Astronomy, University of Southampton, Highfield, Southampton, SO17 1BJ, UK; molinam@astro.soton.ac.uk, ajb@astro.soton.ac.uk, ajd@astro.soton.ac.uk}
\altaffiltext{4}{IASF-Roma/INAF, via del Fosso del Cavaliere 100, I-00133 Roma, Italy; bazzano@rm.iasf.cnr.it, derosa@rm.iasf.cnr.it, ubertini@rm.iasf.cnr.it}
\altaffiltext{5}{CEA Saclay/DSM/DAPNIA/Sap, 91191 Gif Sur Yvette, France; flebrun@cea.fr, fil@discovery.saclay.cea.fr, belanger@hep.saclay.cea.fr}  
\altaffiltext{6}{INTEGRAL Science Data Center, Chemin D'\'Ecogia 16, 1290 Versoix, Switzerland; roland.walter@obs.unige.ch}

\begin{abstract}

Analysis of INTEGRAL Core Program and public Open Time observations performed up 
to April 2005 provides a sample of 62 active galactic nuclei in the 20-100 keV band
above a flux limit of $\sim$ 1.5$\times$10$^{-11}$ erg cm$^{-2}$ s$^{-1}$.
Most(42) of the sources in the sample are Seyfert galaxies, almost equally divided between type 1 and 2 objects, 
6 are blazars and 14 are still unclassified.  Excluding the blazars, the average redshift of our sample is 0.021 
while the mean luminosity is Log(L) = 43.45. We find that absorption is present in 65\% of the objects with 
14\% of the total sample due to Compton thick active galaxies. In agreement with both
Swift/BAT team results and 2-10 keV studies, the 
fraction of absorbed objects decreases with the 20-100 keV luminosity. 
All Seyfert 2s in our sample are absorbed as are 33\% of Seyfert 1s.  
The present data highlight the capability of INTEGRAL to probe the extragalactic gamma-ray sky 
and to find new and/or absorbed active galaxies.
\end{abstract}

\keywords{surveys --- galaxies: active --- gamma rays: observations }

\section{Introduction}

The extragalactic gamma-ray sky is still poorly explored with only a few surveys having been 
performed so far above 10 keV: the all-sky survey conducted in the 1980's  
by the HEAO/A4 instrument in the 13-80 keV band (Levine et al. 1984) and more recently those of 
RXTE/PCA and Swift/BAT, in the 8-20 and 14-195 keV bands respectively (Revnivtsev et al. 2004, 
Markwardt et al. 2005). The latter, characterized by a positional uncertainty of $\leq$3\arcmin 
and a flux detection limit of $\sim$10$^{-11}$ erg cm$^{-2}$ s$^{-1}$, is the most accurate and 
sensitive survey to date at high energies. It covers the high latitude sky providing a sample 
of 44 active galactic nuclei (AGN), most of which are previously known from X-ray studies.

Despite being so rare, gamma-ray surveys are an efficient way to find  AGN as they probe 
heavily obscured regions/objects, i.e. those that could be missed in optical, UV, and even 
X-ray surveys. 
Indeed, 64\% of the non-blazar sources found by Swift/BAT have absorption in excess of 
10$^{22}$ atoms cm$^{-2}$ and the overall column density distribution is bimodal. While none of 
the sources brighter than 3$\times$10$^{43}$ erg s$^{-1}$ shows high column densities, almost 
all weaker objects are absorbed.
Quantifying the fraction of AGN missed by low energy surveys is necessary if we want to provide 
input parameters for synthesis models of the X-ray background and to understand 
the accretion history of the Universe.
 
A further step in this field is provided by the imager (IBIS) on board INTEGRAL, which,
like BAT, is surveying a large fraction of the sky above 20 keV with similar sensitivity 
and positional accuracy. Analysis of the first year of INTEGRAL observations covering largely 
the galactic plane and centre has provided a first sample of 10 gamma-ray selected AGN   
(Bassani et al. 2004); more recently the second IBIS survey has listed 32 such objects 
(Bird et al. 2005). Detailed analysis of INTEGRAL observations
also suggests that AGN are becoming a major constituent of the IBIS 
source population (Revnivtsev et al. 2005, Beckmann et al. 2005); at least 15\% of the objects 
in the second IBIS catalogue are active galaxies. Here, we present a further step
in our all-sky survey project, limiting the search to extragalactic objects.
We have analysed $\sim$11300 INTEGRAL pointings and detected 62 AGN in the 20-100 keV energy 
range. 

\section{Data analysis}

Data reported here belong to the Core Program and public Open Time
observations and span from revolution 46 (February 2003) to revolution
309 (April 2005) inclusive; this represents a significant  
extension both in exposure time and area coverage
with respect to the second IBIS catalogue (Bird et al. 2005) with more than 4000 extra 
pointings being analysed.
A detailed description of the data analysis and source extraction criteria can be found in the
above reference, the only difference being the use of an updated version (4.2) of the standard 
OSA software. 

To search for AGN we have used the 20-100 keV flux map, which provides a good 
combination between significance of detection and overall background level over the mosaic 
image: most AGN have power law spectra with $\Gamma$=1.9 and a break around 100 keV (Malizia et 
al. 2003) and the energy band was selected to match these spectral characteristics. 
The threshold significance level used for the source extraction was
5$\sigma$.

Staring data, which tend to be much noisier than dithering observations, as well as early 
exposures performed while the instrument set-up was still being finalized, were not included in 
the present mosaic; also sources detected only occasionally, e.g. in one or two revolutions
only, are not considered in the present sample although they may be
associated to flaring AGN such as blazars. Due to the different database used, our catalogue 
results may not include some sources already reported in the literature (e.g. Beckmann et al. 
2005). 

For each excess, the flux extracted from the 20-100 keV light curve was then used to 
estimate the source strength (up to 6\% in flux can be lost in the mosaicing process)
and independently confirm the image detection. Once a list of 
reliable excesses was produced, we proceeded to identify them by cross checking the 
IBIS error boxes (assumed to be 3\arcmin as default) with the Simbad, NED (NASA/IPAC Extragalactic Database)
and HEASARC (High Energy Astrophysics Science Archive Research Center) databases.

\section{Soft Gamma-ray selected AGN}

Of all the excesses found in this survey, we report in Table 1 only those which can be 
confidently associated with AGN. 
While some of the other detections may also be active galaxies, due to their location near the 
galactic plane it is difficult to discriminate them from galactic objects without further 
observations.

The 62 sources listed in the table are divided into two sections: (a) 32 objects 
already reported as confirmed or candidate active galaxies in the 2$^{nd}$ IBIS catalogue
(all but one of the candidates have subsequently been confirmed as AGN through follow up 
optical spectroscopy (Masetti et al. 2005a,b)) and (b) 30 new objects, including four which 
were classified as unidentified in the 2$^{nd}$ IBIS survey but which we now consider to be AGN 
candidates.
Between both sets there are 14 sources which have not yet any optical classification, but
their extragalactic nature is strongly indicated either by follow up Chandra observations 
(IGR J07565-4129, IGR J12026-5349 and IGR J17204-3554), 
by their high latitude location (IGR J13000+2529, IGR J13057+2036, IGR J16194-2810 and IGR 
J18429-3243) or by multiwaveband analysis using radio, infrared and X-ray archival data whereby 
their optical counterpart has been found to be associated with a galaxy 
(IGR J07597-3842, IGR J14552-5133, IGR J14492-5535, IGR 
J16558-5293,IGR J20187+4041, IGR J20286+2544 and IGR J21178+5139). 

In the table, we list the relevant IBIS parameters (position, exposure and flux in mCrab)  
together with (where available) the optical 
classification and redshift obtained from NED archive or from more recent publications as 
detailed in the table. Between 20-100 keV, 1 mCrab corresponds, for a Crab like spectrum, to 
1.6$\times$10$^{-11}$ erg cm$^{-2}$s$^{-1}$ which is a value close to our detection limit.
For objects with known distance, fluxes  
have been converted to gamma-ray luminosities assuming H$_0$=71 Km sec$^{-1}$ Mpc$^{-1}$ and 
q$_0$=0 (Spergel et al. 2003). Available column densities are also listed and
have been obtained from X-ray data in the literature or in public databases.
Unfortunately a significant fraction (about 30\%) of our objects
do not have archival X-ray spectra, so that an estimate of the column density must await X-ray 
follow up observations. The quoted column densities are intrinsic to the source, except for 
those cases where N$_{H}$ is comparable to the galactic absorption, 
in which case the quoted values are upper limits.

About 40\% of the objects listed in Table 1 are well known gamma-ray emitters and have
been studied in the IBIS energy range by previous missions
such as BeppoSAX, OSSE and RXTE; comparison of our data with past observations indicates
overall agreement (within a factor of a few) on the flux (Soldi et al. 2005).  
The remaining objects are detected above 10 keV for the first time; many of these
were not even known to be active nor to be X-ray 
sources before their INTEGRAL discovery (Sazonov et al. 2005). This is largely due to their 
location in the galactic plane which has prevented an in-depth 
study of these objects up to now.

A comparison of the IBIS catalogue with the Swift sample indicates that the intersection  
is very small (only 9 sources), due to BAT observing mostly the high 
galactic latitude sky while IBIS maps preferentially the galactic plane.
It is likely that,
together, these two surveys will provide a complete and deep sky coverage and thus the best yet 
sample of gamma-ray selected AGN for some time to come.

\section{Results and Conclusions}

It must be remembered that the present survey is highly inhomogeneous in sky exposure and 
coverage, thus the present sample is far from being  complete. Nevertheless
a few interesting  considerations can be drawn by a simple statistical analysis.
For objects with known distance, we plot in figure 1 the gamma-ray luminosity against redshift,
to show the large range in these parameters sampled by the present survey. From this figure it 
is also evident that our sensitivity limit is around 
1.5$\times$10$^{-11}$ erg cm$^{-2}$s$^{-1}$ (straight line in the figure).

Within the overall sample, 42 objects are classified as Seyfert galaxies, 6 are blazars and 14 
are still unclassified. Within the 
sample of Seyfert galaxies, 23 objects are of type 1-1.5 while 19 are of type 2 thus illustrating the power
of gamma-ray surveys to find narrow line AGN. 
Excluding the blazars, the average redshift of our sample is 
0.021$\pm$0.017 (1$\sigma$) while the mean luminosity in Log is 43.45$\pm$0.71. 
Assuming 10$^{22}$ atoms cm$^{-2}$ to be the dividing line between absorbed and unabsorbed 
sources, we find that absorption is present in 65$\pm$17\% of the sample. This result 
is in line with the Swift findings (Markwardt et al. 2005) and above that found in X-ray 
surveys for bright objects (La Franca et al. 2005, Comastri 2004).
It is also interesting to note that 14$\pm$7\% of the total sample is due to Compton thick 
objects which is about three times the frequency found by Swift. 

In figure 2 we show the column density of the sources as a function of IBIS luminosity. Although from this 
figure there is little evidence of
any strong correlation, the few high luminosity objects in the survey all have low absorption. 
We can further investigate the relationship between
N$_{H}$ and Log(L)  by forming the ratio between the fraction of absorbed sources 
above a given luminosity and that below this value. It can be seen from figure 3 that below 
log(L)$\sim$43.5 the ratio is practically constant while above this luminosity it decreases sharply 
implying that at high IBIS luminosities  there are very few absorbed sources. This is in agreement with 
that found by the Swift/BAT team and obtained from 2-10 keV studies (La Franca et al. 2005).

Within the sub-sample of 17 Seyfert 2s with known N$_{H}$, 
we find that all are absorbed and almost 30\% are Compton thick;
this is in line with previous estimates of the column density distribution of type 2 objects 
based on X-ray data (Risaliti et al. 1999, Bassani et al. 1999). Interestingly we also find 
that 33\% of type 1 objects are absorbed.

More in depth studies of the present sample require optical 
classification of all objects  and a detailed analysis of 
their broad band behaviour (particularly in the X-ray to gamma-ray band) in order to 
understand the role of absorption and the 
relation of extragalactic gamma-ray surveys to those in other wavebands. In the meantime 
this catalogue testifies to the power of INTEGRAL/IBIS to 
probe the extragalactic gamma-ray sky, discover new active galaxies and find
absorbed objects.

\section{Acknowledgements}
This research has been  supported by ASI under
contract I/R/046/04. This research has made use of data obtained from NED (Jet Propulsion
Laboratory, California Institute of Technology), SIMBAD 
(CDS, Strasbourg, France) and HEASARC (NASA's Goddard Space Flight Center).

\begin{deluxetable}{lrrrrlllr}
\tabletypesize{\scriptsize}
\tablecaption{Table 1: Sample of Soft Gamma-ray selected AGN}
\startdata
{\bf Source$^{\#}$}  & {\bf RA}&{\bf Dec}& {\bf Exp(ks)}  &  {\bf {F$^{\dagger}$}}& {\bf Type$^{\star}$}& {\bf z$^{\star}$}&  {\bf N$_{H}$$^{\ddagger}$} & {\bf Log L$^{\dagger}$}\\
\hline
\hline
\multicolumn{9}{c}{AGN in the second IBIS catalogue}\\
\hline
\hline
QSOB0241+62         &  41.210 & +62.481 &  159.9 &  4.3$\pm$0.4    & S1     & 0.044     & 1.5$\pm$0.3$^{(1)}$ & 44.48\\
MCG+8-11-11         &  88.718 & +46.447 &   33.8 &  5.4$\pm$0.8    & S1.5   & 0.020     & $<$0.02$^{(1)}$   & 43.90\\
IGR J07597-3842$^a$ & 119.948 & -38.723 &  892.0 &  2.2$\pm$0.2    & -      & -         &      -              & -    \\
ESO 209-12          & 120.490 & -49.757 & 1062.0 &  1.4$\pm$0.1    & S1.5   & 0.040     &      -              & 43.92\\
Fairall 1146        & 129.649 & -36.023 &  992.0 &  1.0$\pm$0.2    & S1.5   & 0.031     &      -              & 43.54\\
MCG-05-23-016       & 146.895 & -30.932 &  333.6 & 11.2$\pm$1.0    & S2     & 0.008     & 1.6$\pm$0.2$^{(2)}$ & 43.45\\
IGR J10404-4625     & 160.105 & -46.400 &  182.8 &  2.8$\pm$0.4    & S2$^A$ & 0.024$^A$ & $>$1$^{(3)}$        & 43.74\\
NGC4151             & 182.637 & +39.400 &   56.2 & 35.4$\pm$0.5    & S1.5   & 0.003     & 3$\pm$0.4$^{(1)}$   & 43.13\\
4C04.42             & 185.615 &  +4.253 &  208.0 &  2.2$\pm$0.3    & Bl     & 0.965     &   -                 & 47.20\\
NGC4388             & 186.449 & +12.652 &   62.3 & 16.5$\pm$0.7    & S2     & 0.008     & 43$\pm$10$^{(1)}$   & 43.61\\
3C273               & 187.285 &  +2.036 &  270.0 &  8.3$\pm$0.3    & Bl     & 0.158     & $<$0.03$^{(4)}$     & 45.92\\
NGC4507             & 188.904 & -39.904 &  216.0 &  9.4$\pm$0.3    & S2     & 0.012     & 29$\pm$2$^{(1)}$    & 43.66\\
LEDA 170194         & 189.807 & -16.202 &  110.2 &  3.5$\pm$0.5    & S2$^B$ & 0.037$^B$ & 1.9$\pm$0.3$^{(5)}$ & 44.23\\
NGC4593             & 189.917 & -5.349  &  342.0 &  4.4$\pm$0.2    & S1     & 0.009     & $<$0.02$^{(1)}$     & 43.10\\
3C279               & 194.037 & -5.777  &  318.0 &  1.8$\pm$0.2    & Bl     & 0.536     & $<$0.02$^{(4)}$     & 46.46\\
NGC4945             & 196.358 & -49.469 &  470.0 & 16.1$\pm$0.2    & S2     & 0.002     & 400$\pm$80 $^{(1)}$ & 42.29\\
CenA                & 201.363 & -43.021 &  406.0 & 40.8$\pm$0.2    & S2     & 0.002     & 23$\pm$13$^{(2)}$ & 42.67\\
4U1344-60           & 206.882 & -60.619 &  732.0 &  4.2$\pm$0.2    & S1$^A$ & 0.013$^A$ & -                   & 43.39\\
IC4329A             & 207.332 & -30.314 &  236.0 & 12.5$\pm$0.3    & S1.2   & 0.016     & 0.42$\pm$0.02$^{(1)}$& 44.05\\
Circinus            & 213.282 & -65.347 &  644.0 & 12.7$\pm$0.2    & S2     & 0.001     & 360$\pm$70$^{(1)}$  & 41.97\\
IGR J16482-3036     & 252.050 & -30.590 & 1644.0 &  2.0$\pm$0.1    & S1$^A$ & 0.031$^A$ & -                   & 43.83\\
ESO138-G01$^b$      & 253.029 & -59.218 &  572.0 &  1.4$\pm$0.2    & S2     & 0.009     & $>$150$^{(6)}$       & 42.60\\
NGC6300             & 259.234 & -62.816 &  256.0 &  3.8$\pm$0.3    & S2     & 0.004     & 29$\pm$2$^{(7)}$     & 42.25\\
GRS1734-292         & 264.369 & -29.140 & 4040.0 &  5.1$\pm$0.1    & S1     & 0.021     & $<$0.5$^{(8)}$      & 43.92\\
2E1739.1-1210       & 265.466 & -12.199 &  906.0 &  1.5$\pm$0.2    & S1$^C$ & 0.037$^C$ & -                   & 43.87\\
IGRJ18027-1455      & 270.690 & -14.917 & 1476.0 &  2.6$\pm$0.1    & S1     & 0.035     & -                   & 44.06\\
PKS1830-211$^c$     & 278.413 & -21.057 & 1950.0 &  3.1$\pm$0.1$^e$& Bl     & 2.507     & $<$0.26$^{(9)}$     & 48.53\\
ESO103-G35          & 279.695 & -65.408 &   41.7 &  5.3$\pm$0.8    & S2     & 0.013     & 15.1$\pm$0.5$^{(4)}$& 43.51\\
2E1853.7+1534       & 283.984 & +15.610 &  664.0 &  2.0$\pm$0.2    & S1$^B$ & 0.084$^B$ &  -                  & 44.73\\  
NGC6814             & 295.666 & -10.329 &  252.0 &  3.6$\pm$0.3    & S1.5   & 0.005     & $<$0.05$^{(4)}$     & 42.52\\
Cygnus A            & 299.869 & +40.733 &  426.0 &  4.9$\pm$0.2    & S2     & 0.056     & 38$\pm$8$^{(2)}$    & 44.75\\
IGRJ21247+5058      & 321.156 & +50.970 &  438.0 &  6.1$\pm$0.2    & S1     & 0.020     & -                   & 43.93\\
\hline
\hline
\multicolumn{9}{c}{New AGN}\\
\hline
\hline
1ES0033+595         &   9.004 & +59.833 &  776.0 &  0.9$\pm$0.2    & Bl     & 0.086     & 0.36$\pm$0.08$^{(10)}$& 44.38\\
NGC788              &  30.264 &  -6.814 &  134.8 &  4.2$\pm$0.4    & S2     & 0.014     & 21$\pm$0.5$^{(4)}$  & 43.43\\
NGC1068             &  40.704 & -00.007 &  218.0 &  1.5$\pm$0.3    & S2     & 0.004     & $>$1000$^{(1)}$     & 41.88\\
NGC1275$d$          &  49.878 & +41.566 &   78.8 &  3.4$\pm$0.6    & S2     & 0.018     & 1.5$\pm$0.7$^{(2)}$ & 43.57\\
3C111               &  64.611 & +37.998 &   42.4 &  5.5$\pm$0.8    & S1     & 0.049     & $<$0.9$^{(4)}$      & 44.67\\
UGC3142             &  70.988 & +28.960 &   71.7 &  3.9$\pm$0.7    & S1     & 0.022     & -                   & 43.81\\
LEDA168563          &  73.054 & +49.530 &  115.3 &  3.4$\pm$0.5    & S1     & 0.029     & -                   & 44.00\\
MKN3                &  93.854 & +71.044 &  468.0 &  4.8$\pm$0.2    & S2     & 0.014     & 110$\pm$16$^{(1)}$  & 43.49\\
MKN6                & 103.040 & +74.450 &  500.0 &  2.5$\pm$0.2    & S1.5   & 0.019     & 10$\pm$0.6$^{(1)}$  & 43.48\\
IGR J07565-4139$^e$ & 119.080 & -41.613 & 1078.0 &  1.0$\pm$0.1    & -      & -         & 1.1$\pm$0.2$^{(5)}$ & -\\
QSO0836+710         & 130.320 & +70.920 &  412.0 &  3.1$\pm$0.2    & Bl     & 2.172     & $<$0.03$^{(4)}$     & 48.34\\
IGR J12026-5349$^f$ & 180.709 & -53.820 &  334.0 &  2.5$\pm$0.3    & -      & 0.028     & 2.2$\pm$0.3$^{(5)}$ & 43.83\\
IGR J12415-5750$^g$ & 190.368 & -57.851 &  576.0 &  1.1$\pm$0.2    & S2     & 0.024     &  -                  & 43.35\\
IGR J13000+2529$^h$ & 195.022 & +25.490 &  232.0 &  1.3$\pm$0.3    & -      & -         &  -                  & -\\
IGR J13057+2036$^i$ & 196.422 & +20.595 &  174.1 &  1.8$\pm$0.3    & -      & -         &  -                  & -\\
ESO323-G077         & 196.611 & -40.445 &  336.0 &  1.6$\pm$0.2    & S1.2   & 0.015     & 55$\pm$33$^{(11)}$  & 43.11\\
MCG-06-30-15        & 203.976 & -34.297 &  324.0 &  3.1$\pm$0.3    & S1.2   & 0.008     & $<$0.02$^{(12)}$    & 42.81\\
ESO511-G030         & 214.849 & -26.659 &  178.5 &  2.7$\pm$0.4    & S1     & 0.022     & $<$0.05$^{(11)}$    & 43.68\\
IGR J14492-5535$^l$ & 222.318 & -55.587 &  814.0 &  1.2$\pm$0.2    & -      & -         & -                   & -\\
IGR J14552-5133$^m$ & 223.811 & -51.583 &  860.0 &  1.0$\pm$0.2    & -      & -         & -                   & -\\
IGR J16119-6036$^n$ & 242.981 & -60.597 &  600.0 &  1.5$\pm$0.2    & S1     & 0.016     & -                   & 43.11\\
IC4518-A$^o$        & 224.410 & -43.129 &  770.0 &  1.7$\pm$0.2    & S2     & 0.016     & -                   & 43.20\\
IGR J16194-2810$^p$ & 244.874 & -28.110 &  526.0 &  1.7$\pm$0.2    & -      & -         & -                   & -\\
NGC6221$^b$         & 253.029 & -59.218 &  572.0 &  1.4$\pm$0.2    & S2     & 0.005     & 1.1$\pm$0.1$^{(13)}$& 42.07\\
IGR J16558-5203$^q$ & 254.014 & -52.052 &  894.0 &  1.8$\pm$0.2    & -      & -         & -                   & -\\
IGR J17204-3554$^r$ & 260.090 & -35.909 & 3100.0 &  1.3$\pm$0.1    & -      & -         & 14$\pm$1$^{(14)}$   & -\\
IGR J18249-3243$^s$ & 276.245 & -32.715 & 2600.0 &  1.2$\pm$0.1    & -      & -         & -                   & -\\
IGR J20187+4041$^t$ & 304.693 & +40.697 &  576.0 &  1.4$\pm$0.2    & -      & -         & -                   & -\\
IGR J20286+2544$^u$ & 307.156 & +25.765 &  202.0 &  2.3$\pm$0.4    & -      & 0.014     & -                   &43.21\\
IGR J21178+5139$^v$ & 319.429 & +51.671 &  408.0 &  1.6$\pm$0.3    & -      & -         & -                   & -\\
 
\enddata

\tablecomments{
${\dagger}$ Fluxes (in units of mCrab with associated 90\% error) and  luminosities (in  
units of erg s$^{-1}$) in the 20-100 keV band ;
${\ddagger}$ in units of 10$^{22}$ at.cm$^{-2}$\\
${\star}$: Type (S1-1.5= Seyfert 1-1.5; S2=Seyfert2; Bl=Blazar) and redshift according 
to NED or to: (A) Masetti et al. 2005a; (B) Masetti et al. 2005b; (C) Torres et al. 2004\\ 
${\#}$ Notes to source detection: (a) IRAS 07579-3835;
(b) analysis of various energy band mosaics indicates detection of both  
ESO138-G01 and NGC 6221, which are 11' apart; 
(c) lensed galaxy with magnification factor of $\sim$10;
(d) possibly contaminated by the Perseus cluster; 
(e) 2MASX J07561963-4137420, Sazonov et al. 2005; (f) WKK 0560, Sazonov et al. 2005;
(g) WKK1263; (h) possibly MAPS-NGP O-379-0073388; (i) possibly NGP9 F379-1241685; (l) possibly 
2MASX J14491283-5536194; 
(m) WKK4438 ; (n) WKK6092, but contamination by the nearby 
galaxy WKK 6103 is possible; (o) one of a pair of strongly 
interacting galaxies, so far the companion has not been classified as an active galaxy;
(p) 1RXS J161933.6-280736; 
(q) possibly 1RXS J165605.6-520345=IRAS16520-5158;  
(r) NGC 6334B, Bassani et al. 2005;
(s) possibly PKS1821-327;
(t) 2MASX J20183871+4041003;
(u) MCG+04-48-002 possibly interacting with NGC6921;
(v) 2MASX J21174741+5138523.}
\tablerefs{
N$_{H}$ taken from: (1) Lutz et al. 2004; (2) Bassani et al. (1999); 
(3)Masetti et al. (2005a); (4)Tartarus database; 
(5) Sazonov et al. (2005); (6) Collinge \& Brandt 2000; (7) Guainazzi 2002; 
(8) Sazonov et al. (2004); (9) De Rosa et al. (2005), ionized intrinsic absorption also
possible; (10) Donato et al. 2005; (11) Sazonov \& Revnivtsev (2004); (12) Reynolds 
et al. 1997; (13) Panessa \& Bassani 2002; (14) Bassani et al. 2005.}
\end{deluxetable}

\clearpage

\begin{figure}
\plotone{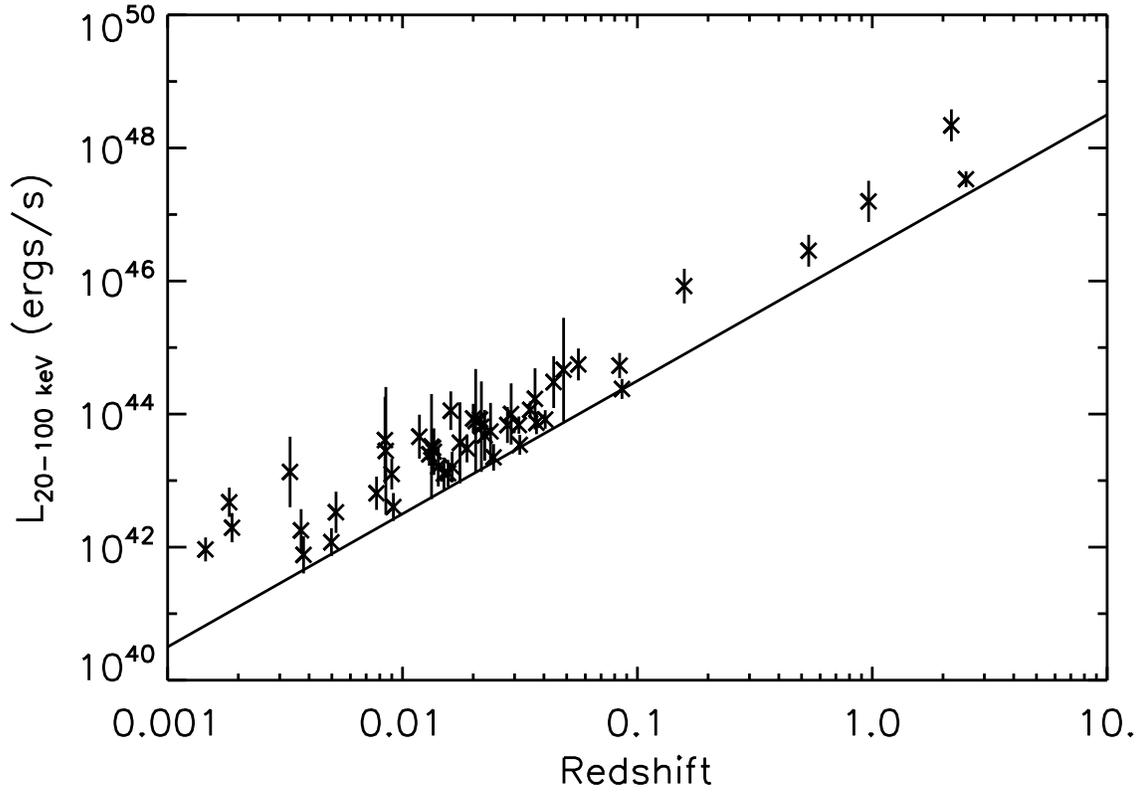}
\caption{20-100 keV luminosity versus redshift for optically classified AGN  in the present survey; 
straight line corresponds to the IBIS survey limit}
\end{figure}

\clearpage

\begin{figure}
\plotone{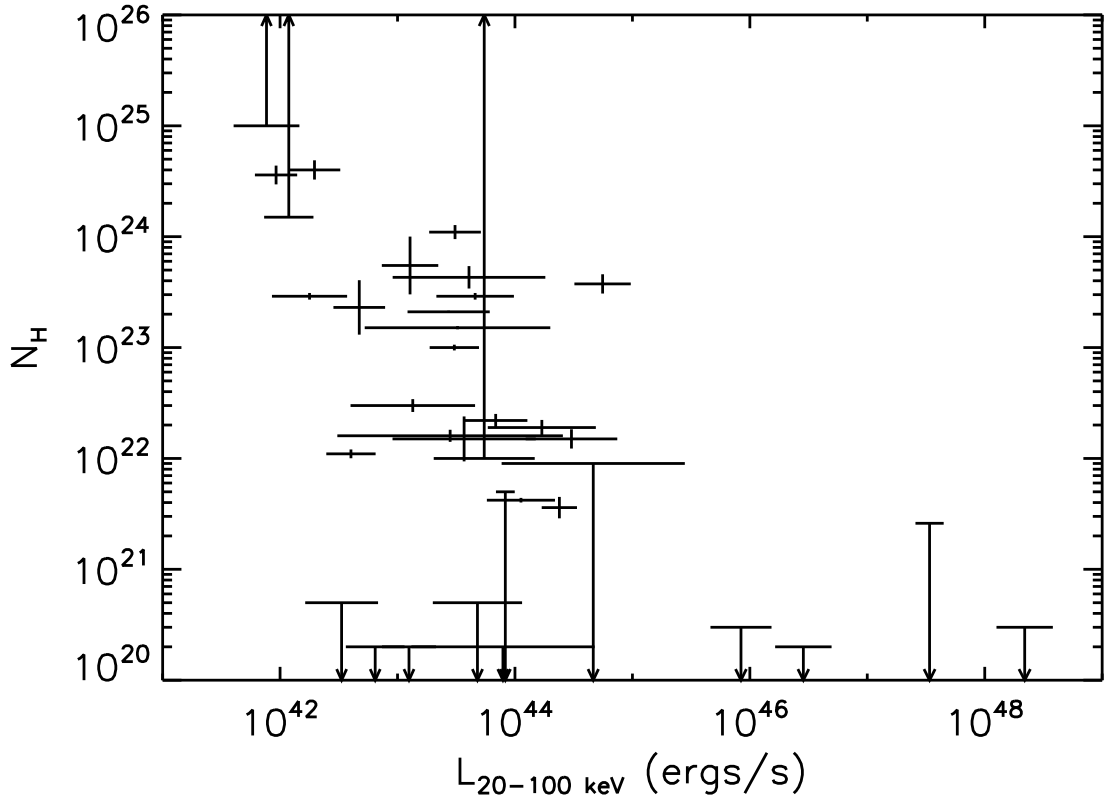}
\caption{Column density versus 20-100 keV luminosity for AGN in the present survey with known intrinsic absorption}
\end{figure}

\clearpage

\begin{figure}
\plotone{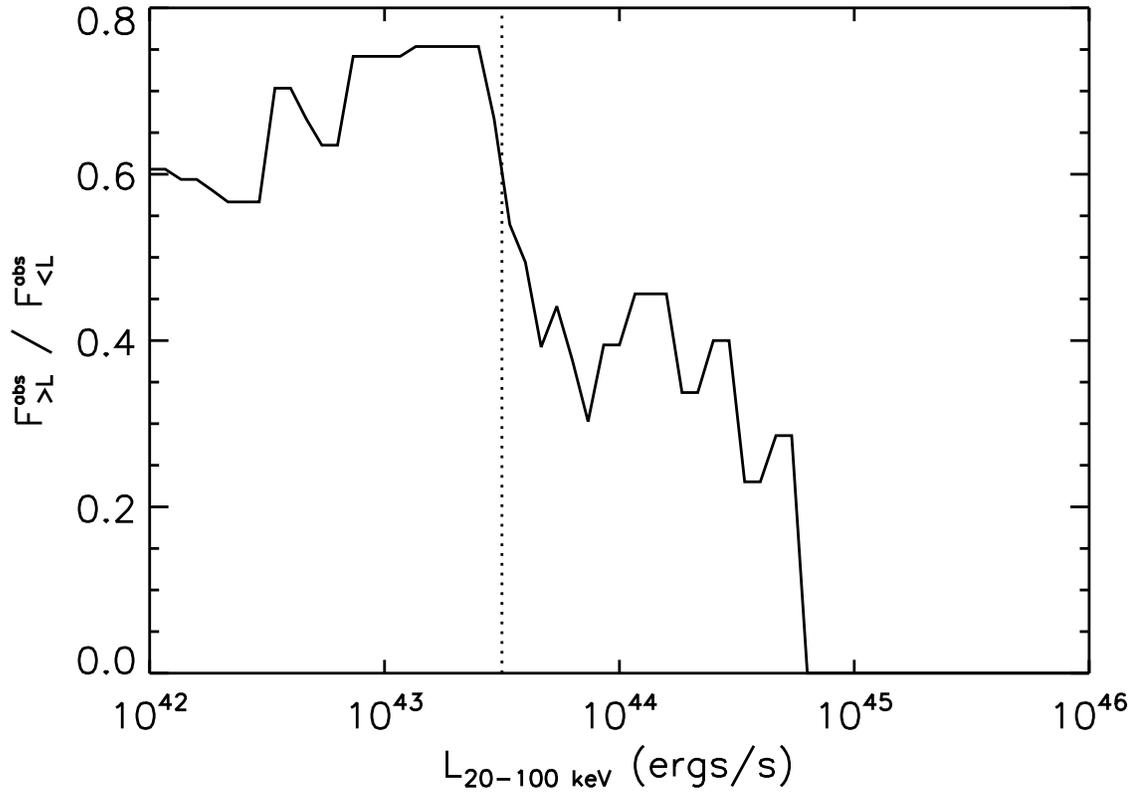}
\caption{Ratio between the fraction of absorbed sources above a given luminosity and that below this value as a function of 20-100 keV luminosity}
\end{figure}

\end{document}